\documentclass[preprint,aps,superscriptaddress,showpacs]{revtex4}
\usepackage[dvips]{graphics} 
\usepackage{graphicx}
\usepackage{longtable}
\usepackage{subfigure}
\usepackage{epsfig}
\usepackage{color}
\begin{document}

\title{Critical Scaling of Two-component Systems from Quantum Fluctuations}
\author{J.~Mabiala}
\email[]{jmabiala@comp.tamu.edu}
\affiliation{Cyclotron Institute, Texas A$\&$M University, College Station, Texas 77843, USA}
\author{A.~Bonasera}
\affiliation{Cyclotron Institute, Texas A$\&$M University, College Station, Texas 77843, USA}
\affiliation{Laboratori Nazionali del Sud, INFN, via Santa Sofia, 62, 95123 Catania, Italy}
\author{H.~Zheng}
\affiliation{Cyclotron Institute, Texas A$\&$M University, College Station, Texas 77843, USA}
\affiliation{Physics Department, Texas A$\&$M University, College Station, Texas 77843, USA}
\author{A.~B.~McIntosh}
\affiliation{Cyclotron Institute, Texas A$\&$M University, College Station, Texas 77843, USA}
\author{Z.~Kohley}
\altaffiliation{Present address: National Superconducting Cyclotron Laboratory, Michigan State University, East Lansing, Michigan 48824, USA}
\affiliation{Cyclotron Institute, Texas A$\&$M University, College Station, Texas 77843, USA}
\affiliation{Chemistry Department, Texas A$\&$M University, College Station, Texas 77843, USA}
\author{P.~Cammarata}
\affiliation{Cyclotron Institute, Texas A$\&$M University, College Station, Texas 77843, USA}
\affiliation{Chemistry Department, Texas A$\&$M University, College Station, Texas 77843, USA}
\author{K.~Hagel}
\affiliation{Cyclotron Institute, Texas A$\&$M University, College Station, Texas 77843, USA}
\author{L.~Heilborn}
\affiliation{Cyclotron Institute, Texas A$\&$M University, College Station, Texas 77843, USA}
\affiliation{Chemistry Department, Texas A$\&$M University, College Station, Texas 77843, USA}
\author{L.~W.~May}
\affiliation{Cyclotron Institute, Texas A$\&$M University, College Station, Texas 77843, USA}
\affiliation{Chemistry Department, Texas A$\&$M University, College Station, Texas 77843, USA}
\author{A.~Raphelt}
\affiliation{Cyclotron Institute, Texas A$\&$M University, College Station, Texas 77843, USA}
\affiliation{Chemistry Department, Texas A$\&$M University, College Station, Texas 77843, USA}
\author{A.~Zarrella}
\affiliation{Cyclotron Institute, Texas A$\&$M University, College Station, Texas 77843, USA}
\affiliation{Chemistry Department, Texas A$\&$M University, College Station, Texas 77843, USA}
\author{S.~J.~Yennello}
\affiliation{Cyclotron Institute, Texas A$\&$M University, College Station, Texas 77843, USA}
\affiliation{Chemistry Department, Texas A$\&$M University, College Station, Texas 77843, USA}

%\date{\today }

\begin{abstract}
The thermodynamics of excited nuclear systems allows one to explore the second-order phase transition in a two-component quantum mixture. Temperatures and densities are derived from quantum fluctuations of fermions. The pressures are determined from the grand partition function of Fisher's model. Critical scaling of observables is found for systems which differ in neutron to proton concentrations thus constraining the equation of state of asymmetric nuclear matter. The derived critical exponent $\beta=0.35 \pm 0.01$, belongs to the liquid-gas universality class. The critical compressibility factor $P_c/\rho_c T_c$ increases with increasing neutron number.
\end{abstract}

\maketitle

Understanding the behavior of nuclear matter under extreme conditions of density and temperature is one way to gain insight into two-component finite quantum systems \cite{Baran, PhysRevC.52.2072}. The basic properties of nuclear matter (consisting of protons,$Z$, and neutrons, $N$) as a function of temperature, density, and proton fraction are described by the nuclear equation of state (EOS). The dependence of the EOS on the proton fraction is essential not only in describing features of heavy-ion reactions but also in describing characteristics of astrophysical environments such as the dynamics of stellar collapse and supernovae as well as the formation and structure of neutron stars \cite{Baran}. Characterization of the critical point of nuclear matter (temperature, density and pressure) as a function of the proton fraction provides important information about the nuclear EOS.

In nuclear matter, the nucleon-nucleon interaction is similar to the van der Waals interaction with short-range repulsion and long-range attraction. This analogy strongly suggests the occurrence of a nuclear phase transition similar to a liquid-gas phase transition \cite{PhysRevC.27.2782, PhysRevLett.75.1040, Laundau.Lifshitz, J.Chem.Phys.13.253, K.Huang}. Experimentally this can be probed with heavy-ion collisions around the Fermi energy. Unlike van der Waals fluids, nuclei are finite, two-component systems. Most of the divergences usually linked to a phase transition in macroscopic systems are washed out in these small systems \cite{Chomaz:2005aa}. Moreover, the additional degree of freedom which is related to proton and neutron concentrations makes the phase transition more complex \cite{PhysRevC.68.014608}. The existence of a nuclear phase transition is currently the subject of much investigation via caloric curves \cite{PhysRevLett.75.1040, PhysRevC.65.034618}, critical exponents \cite{PhysRevLett.49.1321}, negative heat capacities  \cite{D'Agostino:1999kp} and other observables \cite{PhysRevLett.86.3252, Pichon2006267}. However, because of the assumptions made in these studies, the system could not be located in the pressure-density-temperature space \cite{PhysRevC.67.024609}.   

In this paper, we report the experimental quantum temperatures and densities of the fragmenting system calculated by the quantum fluctuation method for protons presented in Refs.~\cite{Zheng:2010kg, Zheng:2011ni}. Since the protons represent the vapor part, the derived densities and temperatures refer to the vapor branch of the `liquid-gas'-like instability region. In addition, pressures calculated through the grand partition function of Fisher's droplet model \cite{Finocchiaro:1995ff} are also presented. It is shown that the present data contain a signature of a liquid-gas phase transition. Scaling of physical observables to their critical values displays universality, i.e. independence from the proton-neutron asymmetry of the source.  In turn this implies that  details of the second-order phase transition do not depend on the nature of the particles (either classical or quantum) nor on their interaction. All systems display universal scaling.  This is quite amazing in view of the fact that quantum methods are used to derive properties of nuclear systems, at variance with the approaches used for classical macroscopic fluids.

The experiment was performed at the Texas A$\&$M University Cyclotron Institute. Beams of $^{64}$Zn, $^{64}$Ni and $^{70}$Zn at 35 MeV/nucleon were incident on targets of $^{64}$Zn, $^{64}$Ni and $^{70}$Zn respectively \cite{KohleyPhD, PhysRevC.83.044601}. The charged particles and free neutrons produced in the reactions were measured with the NIMROD-ISiS 4$\pi$ detector array \cite{Wuenschel2009578}. The granularity and excellent isotopic resolution provided by the array enabled the reconstruction of the quasi-projectile (QP) in mass, charge and excitation energy. The QP is the large excited primary fragment of the projectile following a non-central collision with the target which will subsequently undergo breakup. The Neutron Ball \cite{Schmitt1995487} provided event-by-event experimental information on the free neutrons emitted during a reaction. The QP source was selected by means of event-by-event cuts on the experimental data similar to those used in Refs.~\cite{Wuenschel:2010ix, WuenschelPhD} with its mass ($A=Z+N$) restricted to be in the range $54\leq A \leq 64$. The excitation energy was deduced using the measured free neutron multiplicity, the charged particle kinetic energies, and the energy needed for the breakup (Q-value). 

Recently, the neutron-proton asymmetry of the  source was identified as an additional order parameter in the nuclear phase transition \cite{Bonasera:2008fj, PhysRevC.81.044618, PhysRevC.83.054609}. Therefore, to investigate such a dependence the data were sorted into four different source asymmetry ($m_s=(N-Z)/A$) bins ranging from 0.04 to 0.24 with bin width of 0.05. The mean $m_s$ values corresponding to these $m_s$ bins are 0.065, 0.115, 0.165 and 0.215. The four $m_s$ bins will be referred to with their mean values in the rest of the discussion. The effects of QP excitation energies on the thermodynamic quantities were investigated by further gating the data into nine bins, each 1 MeV wide, in the range of 1-10 MeV/nucleon.

The temperatures of the different selected QPs are calculated with the momentum quadrupole fluctuation method reported in Refs.~\cite{Zheng:2010kg, Zheng:2011ni}. The momentum quadrupole is defined as $Q_{xy} = p^2_x-p^2_y$ where $p_x$ and $p_y$ are the transverse components of a given fragment's momentum. This quantity is zero on average in the center of mass of the equilibrated QP. The longitudinal component, $p_z$, is excluded to minimize contributions from the collision dynamics, which manifest in the beam direction. In this paper, we use protons which are fermions as our probe particle. Assuming a Fermi-Dirac distribution, the normalized transverse fluctuation of the fragment momentum quadrupole ($\sigma_{xy}^2/\bar{N}$) is connected to the temperature $T$ by the relation

\begin{equation}
 \frac{\sigma_{xy}^2}{\bar{N}}=4m^2T^2F_{QC}\ ,
\label{Eq1}
\end{equation}
where $\bar{N}$ and $m$ are the average multiplicity per event and the mass of a given fragment type, respectively.  $F_{QC}$ is the quantum correction factor which should converge to one at high $T$ (classical limit) and is expressed as
\begin{equation}
F_{QC}=0.2\left(\frac{T}{\varepsilon_f}\right)^{-1.71}+1\ ,
\label{Eq2}
\end{equation} 
where $\varepsilon_f = \varepsilon_{f_0}(\rho/\rho_0)^{2/3}$ is the Fermi energy of the nuclear matter at density $\rho$. The quantities $\rho_0$ and $\varepsilon_{f_0}$ denote the normal nuclear density and the corresponding Fermi energy, respectively. The values of $\rho_0=0.15$ fm$^{-3}$ and $\varepsilon_{f_0}$=36 MeV are used in the calculations. 

A similar derivation is given for the normalized multiplicity fluctuation of fermions emitted from the QP, and it is shown to depend on $T/\varepsilon_f$. This quantity in turn is parametrized in terms of $\sigma_N^2/\bar{N}$ which is given as
\begin{eqnarray}
\frac{T}{\varepsilon_f}&=&-0.442+\frac{0.442}{\left(1-\frac{\sigma_N^2}{\bar{N}}\right)^{0.656}}  \nonumber \\ &&+0.345\frac{\sigma_N^2}{\bar{N}}-0.12\left(\frac{\sigma_N^2}{\bar{N}}\right)^2\ .
\label{Eq3}
\end{eqnarray}
Once the normalized fluctuations are experimentally determined for a given excitation energy, using Eq.~\ref{Eq2} and Eq.~\ref{Eq3} the quantum correction factor $F_{QC}$ can be obtained. From Eq.~\ref{Eq1}, one can easily calculate the temperature $T$ and, from the fact that $\varepsilon_f = \varepsilon_{f_0}(\rho/\rho_0)^{2/3}$, the density $\rho$ is obtained. More details can be found in Refs.~\cite{Zheng:2010kg, Zheng:2011ni}.

\begin{figure}[ht]
\centering
\includegraphics[scale=0.85]{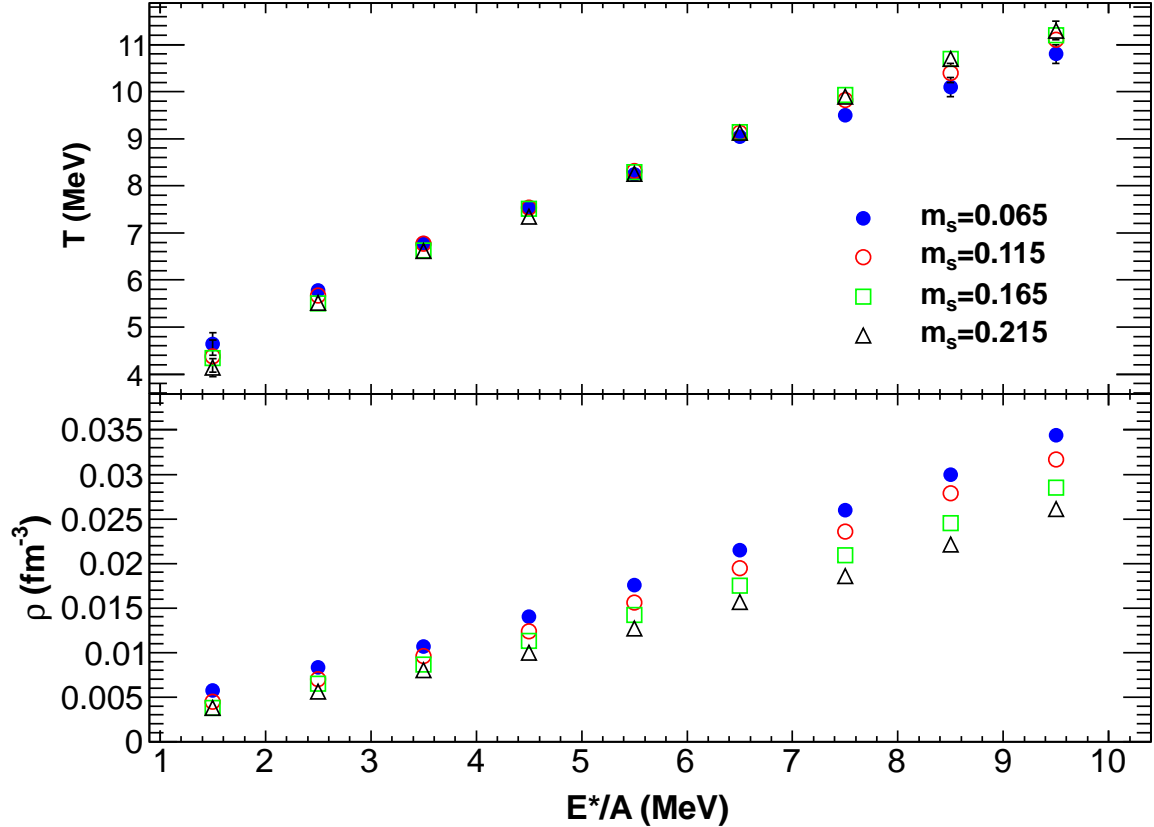}
\caption{(Color online) Temperatures and densities are plotted as a function of the QP excitation energy per nucleon for $54\leq A_{QP}\leq 64$. Both quantities are extracted from the quantum method based on momentum quadrupole and multiplicity fluctuations. Protons are used as the probe particle. Statistical errors are indicated by the bars.}
\label{Fig1}
\end{figure}

Temperatures and densities of the QP are extracted through the momentum quadrupole and multiplicity fluctuations using protons as a probe particle. These are plotted as a function of excitation energy per nucleon in Fig.~\ref{Fig1}. The protons represent the gas or the low-density region in the liquid-gas type phase transition. The extracted densities for the four source asymmetries (bottom panel of Fig.~\ref{Fig1}) show a dependence on the value of $m_s$. An overall ordering in the density with $m_s$ is observed for each excitation energy: the larger the asymmetry, the lower the density. A previous analysis has shown an ordering of the temperatures within a classical treatment \cite{alan}. In the present treatment, the correlation between temperatures and densities indicates that the dependence on $m_s$ is manifest in the densities. However, the spacing between the density values increases as the excitation energy increases. Caloric curves, i.e. temperature as a function of excitation energy (top panel of Fig.~\ref{Fig1}), show a monotonic rising behavior. Within statistical errors, a small dependence on $m_s$ is observed. 

\begin{figure}[ht]
\centering
\includegraphics[width=16cm,height=15cm,angle=0]{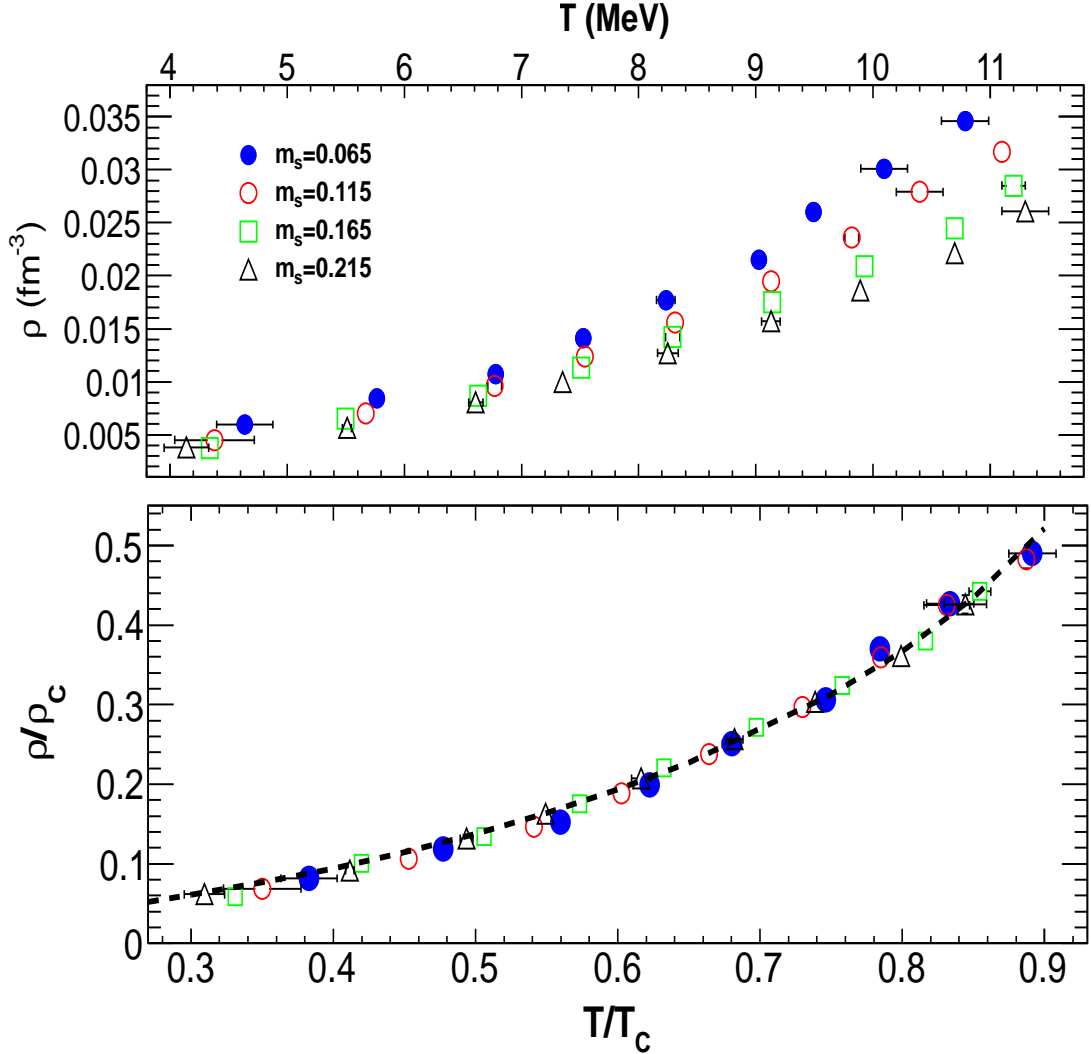}
\caption{(Color online) Top panel: Density plotted as a function of temperature. Bottom panel: The reduced density-temperature phase diagram. The dashed line is drawn to guide the eye. Statistical errors are indicated by the bars.}
\label{Fig2}
\end{figure}

The density $\rho$ is plotted as a function of the temperature $T$ in the top panel of Fig.~\ref{Fig2} for the four different source asymmetries. Universality of a second-order phase transition dictates that the different curves in Fig.~\ref{Fig2} should collapse to one when the axes are divided by their corresponding critical values for each $m_s$.  In turn, we can determine the values of $\rho_c$ and $T_c$ for each $m_s$ in order to obtain one universal curve. In the bottom part of Fig.~\ref{Fig2} we demonstrate that indeed the data displays universality.

\begin{figure}[ht]
\centering
\includegraphics[width=16cm,height=14cm,angle=0]{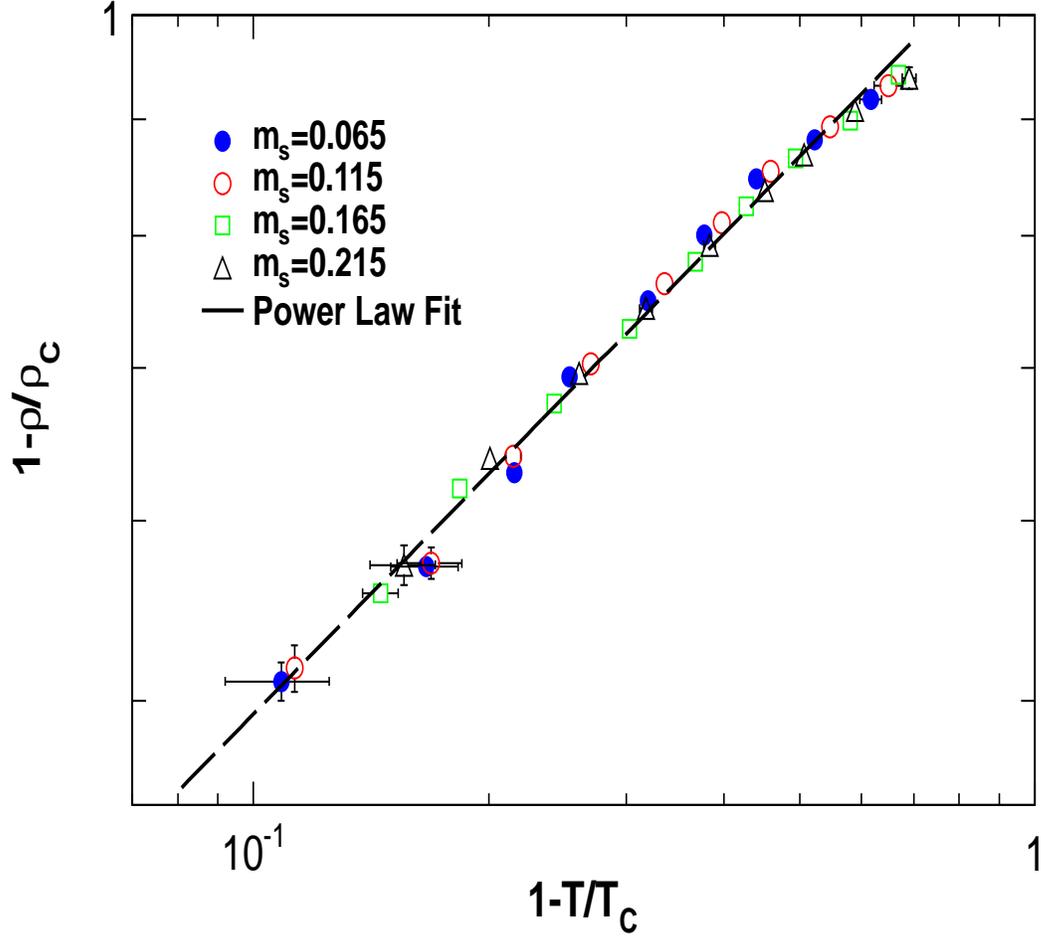}
\caption{(Color online) The extraction of the critical exponent $\beta$, Eq.~\ref{Eq4}. See text for details.}
\label{Fig3}
\end{figure}

The critical exponent $\beta$ is determined from the relation \cite{K.Huang}
\begin{equation}
1-\frac{\rho}{\rho_c}\propto \left(1-\frac{T}{T_c}\right)^{\beta}\ . 
\label{Eq4}
\end{equation}
This parameter defines the universality class of the system and, therefore, systems with similar $\beta$ values have similar underlying physics. Figure \ref{Fig3} shows an excellent fit to the data points that results in a slope $\beta=0.35 \pm 0.01$. This value is in the range of that expected for the liquid-gas universality class \cite{K.Huang, PhysRevLett.88.042701}. The fact that we are getting a value of $\beta$ consistent with a liquid-gas phase transition supports our strategy for calculating densities and temperatures from the Fermi gas assumption.

\begin{figure}[!ht]
\centering
\includegraphics[width=16cm,height=12cm,angle=0]{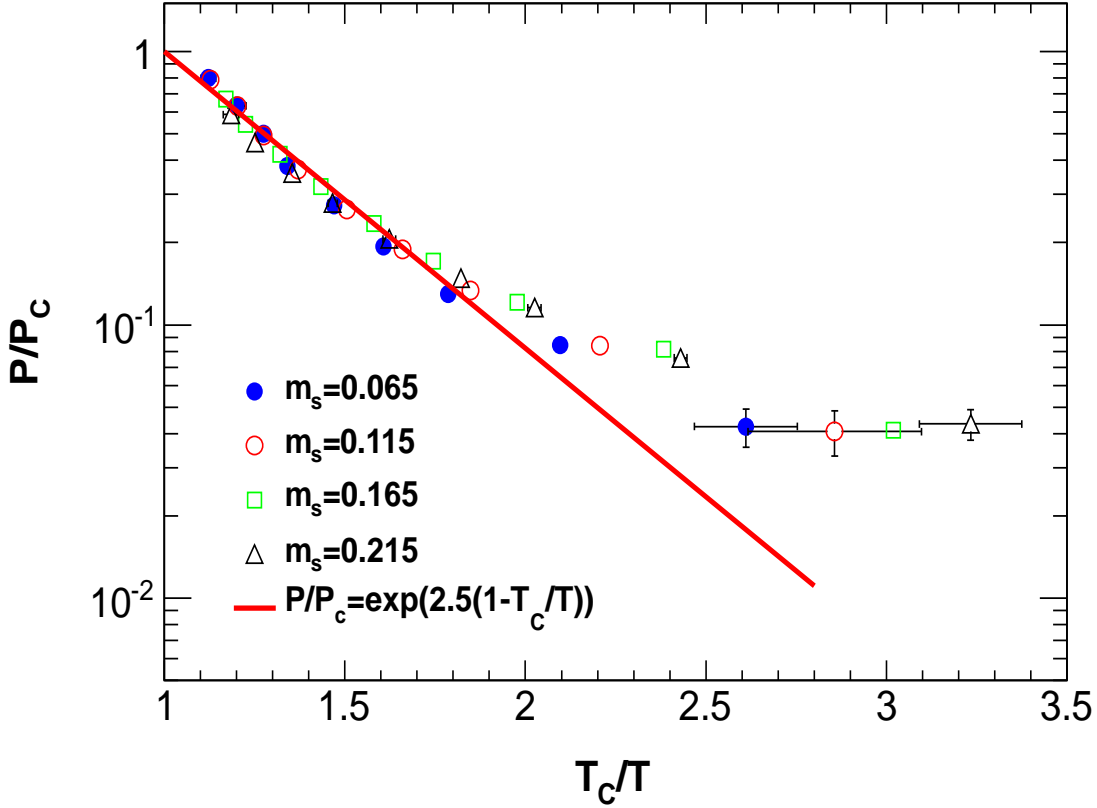}
\caption{(Color online) The reduced pressure as a function of the inverse of the reduced temperature (symbols with statistical error bars) as derived by Guggenheim \cite{Guggenheim} according to the principle of corresponding states. The solid line shows a fit to the Clausius-Clapeyron equation where the value of 2.5 is the average of the $\Delta H/T_c$ values extracted from different $m_s$ bins.}
\label{Fig4}
\end{figure}

The pressure of the system is calculated by making use of the grand partition function from Fisher's droplet model which is based on the simple idea that a real gas of interacting particles can be considered as an ideal gas of clusters (fragments) of various sizes in chemical equilibrium \cite{Finocchiaro:1995ff}. The two-body interaction is assumed to be exhausted in the formation of clusters and the resulting clusters behave ideally. The fragment size fluctuations are strongly enhanced for systems undergoing a phase transition. The equation of state is expressed in terms of the zeroth ($M_0$) and first ($M_1$) moments of the fragment (cluster) size distribution by
\begin{equation}
P=T\rho\frac{M_0}{M_1}\ ,
\label{Eq5}
\end{equation}
where $P$ is the pressure. The temperature and density have been determined from the proton quantum fluctuations and can be used in Eq.~\ref{Eq5}. The $k$-th moment is defined as
\begin{equation}
M_k=\sum_{A\neq A_{max}}A^kY(A)\ ,
\end{equation}
where $Y(A)$ is the multiplicity of the fragment $A$ and the largest fragment $A_{max}$ which represents the liquid phase is excluded in the summation. The calculated pressure $P$ is normalized to its critical value  $P_c$ and plotted versus $T_c/T$, the inverse of the reduced temperature in Fig.~\ref{Fig4}. Even though data points from the lowest two excitation energy bins deviate from the fitted curve, it is interesting to notice that an excellent fit was obtained in the region close to the critical values. The critical pressure is determined using the equation $\ln P/P_c=A-BT_c/T$, with $A$ and $B$ as constants, derived by Guggenheim from the principle of corresponding states \cite{Guggenheim}. Taking $A$ to be nearly equal to $B$, as it is for van der Waals systems, one obtains $P/P_c=\exp[\Delta H/T_c(1-T_c/T)]$ which is the Clausius-Clapeyron equation that describes several fluids for $T\leq T_c$. The quantity $\Delta H$ is the enthalpy of emission of a fragment from the liquid. 

\begin{table*}[tbp]
%\centering
%\begin{table*}[tbp]
\caption{Critical values and thermodynamic quantities for the four $m_s$ bins. }
\label{critical_values}%
\begin{ruledtabular}
\begin{tabular}{cccccc}
%\hline\hline
$m_s$ & $T_c$ (MeV) & $\rho_c$ (fm$^{-3}$) & $P_c$ (MeV/fm$^3$) & $P_c/ \rho_cT_c$ & $\Delta H$ (MeV) \\ \hline
0.065 & 12.12 $\pm$ 0.39 & 0.070 $\pm$ 0.006 & 0.211 $\pm$ 0.002 & 0.25 $\pm$ 0.02 & 31.50 $\pm$ 1.01 \\ 
0.115 & 12.51 $\pm$ 0.35 & 0.066 $\pm$ 0.005 & 0.209 $\pm$ 0.001 & 0.25 $\pm$ 0.02 & 32.53 $\pm$ 0.90 \\  
0.165 &  13.11 $\pm$ 0.30 & 0.064 $\pm$ 0.004 & 0.232 $\pm$ 0.001 & 0.27 $\pm$ 0.02 & 31.46 $\pm$ 0.71 \\ 
0.215 & 13.39 $\pm$ 0.21 & 0.061 $\pm$ 0.002 & 0.258 $\pm$ 0.002 & 0.31 $\pm$ 0.01 & 32.13 $\pm$ 0.50\\ 
%\hline\hline
\end{tabular}
\end{ruledtabular}
\end{table*}

The experimental critical parameters ($T_c$, $\rho_c$ and $P_c$) that give the location of the critical point along with the critical compressibility factor for each $m_s$ bin are listed in Table~\ref{critical_values}. The critical temperatures $T_c$ are observed to increase when increasing $m_s$. This trend is consistent with the results reported in Ref.~\cite{Bonasera:2008fj}. On the other hand, the critical densities $\rho_c$ decrease with increasing $m_s$ while the critical pressures $P_c$ generally increase with $m_s$. The critical compressibility factor values ($P_c/\rho_c T_c$), which quantify the deviation from ideal gas, in turn increase when increasing $m_s$. These values are very close to those of real gases \cite{PhysRevLett.24.47, J.Chem.Phys.13.253}. The values of $\Delta H$ are also reported in Table~\ref{critical_values} and are observed to   weakly depend on $m_s$. 

%% Conclusion
\begin{sloppypar}
By means of a new quantum method and the grand partition function of Fisher's droplet model, the temperature, the density and the pressure of the selected fragmenting sources have been calculated and the phase diagrams have been constructed. These parameters and the corresponding critical values have shown a dependence on the source neutron/proton concentration. A complete experimental location of the critical point is given. The critical exponent $\beta$ and the critical compressibility factor have been extracted and found to belong to the liquid-gas universality class. Strong evidence for a signature of a liquid-gas phase transition in two-component systems has been found. These results provide a means to establish the proton-fraction dependence of the EOS in systems with large neutron excess such as neutron stars. 
\end{sloppypar}

\begin{acknowledgments}
We are indebted to the staff of the TAMU Cyclotron Institute for providing the high quality beams during the experimental work. This work was performed with financial support from the Robert A. Welch Foundation (A-1266), and the U. S. Department of Energy (DE-FG03-93ER-40773).
\end{acknowledgments}

%\nocite{*}
%\bibliographystyle{apsrev}
%\bibliographystyle{plain}
\bibliographystyle{h-physrev3.bst}
%\bibliographystyle{apsrev}
%\bibliography{acompat,biblio}
\bibliography{reference}

\begin{thebibliography}{31}
\expandafter\ifx\csname natexlab\endcsname\relax\def\natexlab#1{#1}\fi
\expandafter\ifx\csname bibnamefont\endcsname\relax
  \def\bibnamefont#1{#1}\fi
\expandafter\ifx\csname bibfnamefont\endcsname\relax
  \def\bibfnamefont#1{#1}\fi
\expandafter\ifx\csname citenamefont\endcsname\relax
  \def\citenamefont#1{#1}\fi
\expandafter\ifx\csname url\endcsname\relax
  \def\url#1{\texttt{#1}}\fi
\expandafter\ifx\csname urlprefix\endcsname\relax\def\urlprefix{URL }\fi
\providecommand{\bibinfo}[2]{#2}
\providecommand{\eprint}[2][]{\url{#2}}

\bibitem[{\citenamefont{Baran et~al.}(2005)\citenamefont{Baran, Colonna, Greco,
  and Di~Toro}}]{Baran}
\bibinfo{author}{\bibfnamefont{V.}~\bibnamefont{Baran}},
  \bibinfo{author}{\bibfnamefont{M.}~\bibnamefont{Colonna}},
  \bibinfo{author}{\bibfnamefont{V.}~\bibnamefont{Greco}}, \bibnamefont{and}
  \bibinfo{author}{\bibfnamefont{M.}~\bibnamefont{Di~Toro}},
  \bibinfo{journal}{Phys. Rept.} \textbf{\bibinfo{volume}{410}},
  \bibinfo{pages}{335} (\bibinfo{year}{2005}).

\bibitem[{\citenamefont{M\"uller and Serot}(1995)}]{PhysRevC.52.2072}
\bibinfo{author}{\bibfnamefont{H.}~\bibnamefont{M\"uller}} \bibnamefont{and}
  \bibinfo{author}{\bibfnamefont{B.~D.} \bibnamefont{Serot}},
  \bibinfo{journal}{Phys. Rev. C} \textbf{\bibinfo{volume}{52}},
  \bibinfo{pages}{2072} (\bibinfo{year}{1995}).

\bibitem[{\citenamefont{Jaqaman et~al.}(1983)\citenamefont{Jaqaman, Mekjian,
  and Zamick}}]{PhysRevC.27.2782}
\bibinfo{author}{\bibfnamefont{H.}~\bibnamefont{Jaqaman}},
  \bibinfo{author}{\bibfnamefont{A.~Z.} \bibnamefont{Mekjian}},
  \bibnamefont{and} \bibinfo{author}{\bibfnamefont{L.}~\bibnamefont{Zamick}},
  \bibinfo{journal}{Phys. Rev. C} \textbf{\bibinfo{volume}{27}},
  \bibinfo{pages}{2782} (\bibinfo{year}{1983}).

\bibitem[{\citenamefont{Pochodzalla et~al.}(1995)\citenamefont{Pochodzalla,
  et~al.}}]{PhysRevLett.75.1040}
\bibinfo{author}{\bibfnamefont{J.}~\bibnamefont{Pochodzalla}}, ,
  \bibnamefont{et~al.}, \bibinfo{journal}{Phys. Rev. Lett.}
  \textbf{\bibinfo{volume}{75}}, \bibinfo{pages}{1040} (\bibinfo{year}{1995}).

\bibitem[{\citenamefont{Landau and Lifshitz}(1980)}]{Laundau.Lifshitz}
\bibinfo{author}{\bibfnamefont{L.~D.} \bibnamefont{Landau}} \bibnamefont{and}
  \bibinfo{author}{\bibfnamefont{E.~M.} \bibnamefont{Lifshitz}},
  \emph{\bibinfo{title}{Statistical Physics}} (\bibinfo{publisher}{Pergamon,
  New York}, \bibinfo{year}{1980}).

\bibitem[{\citenamefont{Guggenheim}(1945)}]{J.Chem.Phys.13.253}
\bibinfo{author}{\bibfnamefont{E.~A.} \bibnamefont{Guggenheim}},
  \bibinfo{journal}{J. Chem. Phys.} \textbf{\bibinfo{volume}{13}},
  \bibinfo{pages}{253} (\bibinfo{year}{1945}).

\bibitem[{\citenamefont{Huang}(1987)}]{K.Huang}
\bibinfo{author}{\bibfnamefont{K.}~\bibnamefont{Huang}},
  \emph{\bibinfo{title}{Statistical Mechanics}} (\bibinfo{publisher}{Wiley $\&$
  Sons, New York}, \bibinfo{year}{1987}).

\bibitem[{\citenamefont{Chomaz and Gulminelli}(2005)}]{Chomaz:2005aa}
\bibinfo{author}{\bibfnamefont{P.}~\bibnamefont{Chomaz}} \bibnamefont{and}
  \bibinfo{author}{\bibfnamefont{F.}~\bibnamefont{Gulminelli}},
  \bibinfo{journal}{Nucl. Phys. A} \textbf{\bibinfo{volume}{749}},
  \bibinfo{pages}{3c} (\bibinfo{year}{2005}).

\bibitem[{\citenamefont{Lee and Mekjian}(2003)}]{PhysRevC.68.014608}
\bibinfo{author}{\bibfnamefont{S.~J.} \bibnamefont{Lee}} \bibnamefont{and}
  \bibinfo{author}{\bibfnamefont{A.~Z.} \bibnamefont{Mekjian}},
  \bibinfo{journal}{Phys. Rev. C} \textbf{\bibinfo{volume}{68}},
  \bibinfo{pages}{014608} (\bibinfo{year}{2003}).

\bibitem[{\citenamefont{Natowitz et~al.}(2002)}]{PhysRevC.65.034618}
\bibinfo{author}{\bibfnamefont{J.~B.} \bibnamefont{Natowitz}}
  \bibnamefont{et~al.}, \bibinfo{journal}{Phys. Rev. C}
  \textbf{\bibinfo{volume}{65}}, \bibinfo{pages}{034618}
  (\bibinfo{year}{2002}).

\bibitem[{\citenamefont{Finn et~al.}(1982)}]{PhysRevLett.49.1321}
\bibinfo{author}{\bibfnamefont{J.~E.} \bibnamefont{Finn}} \bibnamefont{et~al.},
  \bibinfo{journal}{Phys. Rev. Lett.} \textbf{\bibinfo{volume}{49}},
  \bibinfo{pages}{1321} (\bibinfo{year}{1982}).

\bibitem[{\citenamefont{D'Agostino et~al.}(2000)\citenamefont{D'Agostino,
  Gulminelli, Chomaz, Bruno, Cannata et~al.}}]{D'Agostino:1999kp}
\bibinfo{author}{\bibfnamefont{M.}~\bibnamefont{D'Agostino}},
  \bibinfo{author}{\bibfnamefont{F.}~\bibnamefont{Gulminelli}},
  \bibinfo{author}{\bibfnamefont{P.}~\bibnamefont{Chomaz}},
  \bibinfo{author}{\bibfnamefont{M.}~\bibnamefont{Bruno}},
  \bibinfo{author}{\bibfnamefont{F.}~\bibnamefont{Cannata}},
  \bibnamefont{et~al.}, \bibinfo{journal}{Phys. Lett. B}
  \textbf{\bibinfo{volume}{473}}, \bibinfo{pages}{219} (\bibinfo{year}{2000}).

\bibitem[{\citenamefont{Borderie et~al.}(2001)}]{PhysRevLett.86.3252}
\bibinfo{author}{\bibfnamefont{B.}~\bibnamefont{Borderie}}
  \bibnamefont{et~al.}, \bibinfo{journal}{Phys. Rev. Lett.}
  \textbf{\bibinfo{volume}{86}}, \bibinfo{pages}{3252} (\bibinfo{year}{2001}).

\bibitem[{\citenamefont{Pichon et~al.}(2006)}]{Pichon2006267}
\bibinfo{author}{\bibfnamefont{M.}~\bibnamefont{Pichon}} \bibnamefont{et~al.},
  \bibinfo{journal}{Nucl. Phys. A} \textbf{\bibinfo{volume}{779}},
  \bibinfo{pages}{267 } (\bibinfo{year}{2006}).

\bibitem[{\citenamefont{Elliott et~al.}(2003)}]{PhysRevC.67.024609}
\bibinfo{author}{\bibfnamefont{J.~B.} \bibnamefont{Elliott}}
  \bibnamefont{et~al.}, \bibinfo{journal}{Phys. Rev. C}
  \textbf{\bibinfo{volume}{67}}, \bibinfo{pages}{024609}
  (\bibinfo{year}{2003}).

\bibitem[{\citenamefont{Zheng and Bonasera}(2011{\natexlab{a}})}]{Zheng:2010kg}
\bibinfo{author}{\bibfnamefont{H.}~\bibnamefont{Zheng}} \bibnamefont{and}
  \bibinfo{author}{\bibfnamefont{A.}~\bibnamefont{Bonasera}},
  \bibinfo{journal}{Phys. Lett. B} \textbf{\bibinfo{volume}{696}},
  \bibinfo{pages}{178} (\bibinfo{year}{2011}{\natexlab{a}}).

\bibitem[{\citenamefont{Zheng and Bonasera}(2011{\natexlab{b}})}]{Zheng:2011ni}
\bibinfo{author}{\bibfnamefont{H.}~\bibnamefont{Zheng}} \bibnamefont{and}
  \bibinfo{author}{\bibfnamefont{A.}~\bibnamefont{Bonasera}},
  \bibinfo{journal}{nucl-th/1112.4098}  (\bibinfo{year}{2011}{\natexlab{b}}).

\bibitem[{\citenamefont{Finocchiaro et~al.}(1996)\citenamefont{Finocchiaro,
  Belkacem, Kubo, Latora, and Bonasera}}]{Finocchiaro:1995ff}
\bibinfo{author}{\bibfnamefont{P.}~\bibnamefont{Finocchiaro}},
  \bibinfo{author}{\bibfnamefont{M.}~\bibnamefont{Belkacem}},
  \bibinfo{author}{\bibfnamefont{T.}~\bibnamefont{Kubo}},
  \bibinfo{author}{\bibfnamefont{V.}~\bibnamefont{Latora}}, \bibnamefont{and}
  \bibinfo{author}{\bibfnamefont{A.}~\bibnamefont{Bonasera}},
  \bibinfo{journal}{Nucl. Phys. A} \textbf{\bibinfo{volume}{600}},
  \bibinfo{pages}{236} (\bibinfo{year}{1996}).

\bibitem[{\citenamefont{Kohley}(2010)}]{KohleyPhD}
\bibinfo{author}{\bibfnamefont{Z.}~\bibnamefont{Kohley}}, Ph.D. thesis,
  \bibinfo{school}{Texas A$\&$M University} (\bibinfo{year}{2010}).

\bibitem[{\citenamefont{Kohley et~al.}(2011)}]{PhysRevC.83.044601}
\bibinfo{author}{\bibfnamefont{Z.}~\bibnamefont{Kohley}} \bibnamefont{et~al.},
  \bibinfo{journal}{Phys. Rev. C} \textbf{\bibinfo{volume}{83}},
  \bibinfo{pages}{044601} (\bibinfo{year}{2011}).

\bibitem[{\citenamefont{Wuenschel et~al.}(2009)}]{Wuenschel2009578}
\bibinfo{author}{\bibfnamefont{S.}~\bibnamefont{Wuenschel}}
  \bibnamefont{et~al.}, \bibinfo{journal}{Nucl. Instrum. Methods A}
  \textbf{\bibinfo{volume}{604}}, \bibinfo{pages}{578} (\bibinfo{year}{2009}).

\bibitem[{\citenamefont{Schmitt et~al.}(1995)}]{Schmitt1995487}
\bibinfo{author}{\bibfnamefont{R.}~\bibnamefont{Schmitt}} \bibnamefont{et~al.},
  \bibinfo{journal}{Nucl. Instrum. Methods A} \textbf{\bibinfo{volume}{354}},
  \bibinfo{pages}{487 } (\bibinfo{year}{1995}).

\bibitem[{\citenamefont{Wuenschel et~al.}(2010)}]{Wuenschel:2010ix}
\bibinfo{author}{\bibfnamefont{S.}~\bibnamefont{Wuenschel}}
  \bibnamefont{et~al.}, \bibinfo{journal}{Nucl. Phys. A}
  \textbf{\bibinfo{volume}{843}}, \bibinfo{pages}{1} (\bibinfo{year}{2010}).

\bibitem[{\citenamefont{Wuenschel}(2009)}]{WuenschelPhD}
\bibinfo{author}{\bibfnamefont{S.}~\bibnamefont{Wuenschel}}, Ph.D. thesis,
  \bibinfo{school}{Texas A$\&$M University} (\bibinfo{year}{2009}).

\bibitem[{\citenamefont{Bonasera et~al.}(2008)\citenamefont{Bonasera, Chen,
  Wada, Hagel, Natowitz et~al.}}]{Bonasera:2008fj}
\bibinfo{author}{\bibfnamefont{A.}~\bibnamefont{Bonasera}},
  \bibinfo{author}{\bibfnamefont{Z.}~\bibnamefont{Chen}},
  \bibinfo{author}{\bibfnamefont{R.}~\bibnamefont{Wada}},
  \bibinfo{author}{\bibfnamefont{K.}~\bibnamefont{Hagel}},
  \bibinfo{author}{\bibfnamefont{J.}~\bibnamefont{Natowitz}},
  \bibnamefont{et~al.}, \bibinfo{journal}{Phys. Rev. Lett.}
  \textbf{\bibinfo{volume}{101}}, \bibinfo{pages}{122702}
  (\bibinfo{year}{2008}).

\bibitem[{\citenamefont{Huang et~al.}(2010)}]{PhysRevC.81.044618}
\bibinfo{author}{\bibfnamefont{M.}~\bibnamefont{Huang}} \bibnamefont{et~al.},
  \bibinfo{journal}{Phys. Rev. C} \textbf{\bibinfo{volume}{81}},
  \bibinfo{pages}{044618} (\bibinfo{year}{2010}).

\bibitem[{\citenamefont{Tripathi et~al.}(2011)}]{PhysRevC.83.054609}
\bibinfo{author}{\bibfnamefont{R.}~\bibnamefont{Tripathi}}
  \bibnamefont{et~al.}, \bibinfo{journal}{Phys. Rev. C}
  \textbf{\bibinfo{volume}{83}}, \bibinfo{pages}{054609}
  (\bibinfo{year}{2011}).

\bibitem[{\citenamefont{McIntosh et~al.}()}]{alan}
\bibinfo{author}{\bibfnamefont{A.~B.} \bibnamefont{McIntosh}}
  \bibnamefont{et~al.}, \bibinfo{journal}{submitted to Phys. Rev. Lett}
  (????).

\bibitem[{\citenamefont{Elliott et~al.}(2002)}]{PhysRevLett.88.042701}
\bibinfo{author}{\bibfnamefont{J.~B.} \bibnamefont{Elliott}}
  \bibnamefont{et~al.}, \bibinfo{journal}{Phys. Rev. Lett.}
  \textbf{\bibinfo{volume}{88}}, \bibinfo{pages}{042701}
  (\bibinfo{year}{2002}).

\bibitem[{\citenamefont{Guggenheim}(1993)}]{Guggenheim}
\bibinfo{author}{\bibfnamefont{E.~A.} \bibnamefont{Guggenheim}},
  \emph{\bibinfo{title}{Thermodynamics}} (\bibinfo{publisher}{North-Holland,
  Amsterdam}, \bibinfo{year}{1993}).

\bibitem[{\citenamefont{Kiang}(1970)}]{PhysRevLett.24.47}
\bibinfo{author}{\bibfnamefont{C.~S.} \bibnamefont{Kiang}},
  \bibinfo{journal}{Phys. Rev. Lett.} \textbf{\bibinfo{volume}{24}},
  \bibinfo{pages}{47} (\bibinfo{year}{1970}).

\end{thebibliography}

\end{document}